The A-B transition in superfluid $^3$He under confinement in a thin slab geometry

N. Zhelev[1], T.S. Abhilash[1], E.N. Smith[1], R.G. Bennett[1],

X. Rojas[2], L. Levitin[2], J. Saunders[2], and J.M. Parpia[1]

[1]Department of Physics, Cornell University, Ithaca, NY, 14853 USA

[2]Department of Physics, Royal Holloway University of London, Egham, TW20 0EX Surrey,

United Kingdom

**Abstract:**

**The influence of confinement on the topological phases of superfluid $^3$He is studied using the torsional pendulum method. We focus on the phase transition between the chiral A-phase and the time-reversal-invariant B-phase, motivated by the prediction of a spatially-modulated (stripe) phase at the A-B phase boundary. We confine superfluid $^3$He to a single 1.08 μm thick nanofluidic cavity incorporated into a high-precision torsion pendulum, and map the phase diagram between 0.1 and 5.6 bar. We observe only small supercooling of the A-phase, in comparison to bulk or when confined in aerogel. This has a non-monotonic pressure dependence, suggesting that a new intrinsic B-phase nucleation mechanism operates under confinement, mediated by the putative stripe phase. Both the phase diagram and the relative superfluid fraction of the A and B phases, show that strong coupling is present at all pressures, with implications for the stability of the stripe phase.**

**Introduction:**

Superfluid ³He is one of the richest condensed matter systems. Its impact extends to fields as diverse as unconventional superconductivity[1-5], cosmology[6-10] and turbulence[11]. The ³He *p-wave* superfluid order parameter is described by a 3×3 matrix encoding the orientation of the *s*=1 spin and *l* = 1 orbital angular momentum vectors over the Fermi surface. In zero magnetic field two superfluid phases A and B, emerge to break the rotational symmetry of the normal state in different ways.[12,13] These superfluids belong to important classes of topological quantum matter and serve as model systems for topological superconductivity. [14-16]

A new approach in the study of topological superfluidity is to confine the ³He in a precisely engineered nanofabricated geometry.[17] Here we report on a study using a torsion pendulum to investigate the influence of confinement on superfluid ³He within a simple, regular well characterized cavity geometry with height of order the Cooper pair diameter, $\xi_0$. The torsion pendulum measures the superfluid density, and examines the profound influence of confinement on the A-B phase boundary in zero magnetic field, and the nucleation of B phase from the A phase.

We make the striking observation that the supercooling of the first order A-B transition is very small, far less than the considerable supercooling observed in bulk[18], or in anisotropic aerogel[19,20] and discuss the new insights this brings to the mystery of B-phase nucleation. This is of interest not only as a transition between two different topological phases, but also between two quantum vacua of different symmetry, with cosmological analogues to symmetry breaking phase transitions in the early universe responsible for its large scale structure and the dominance of matter over anti-matter.[6-11,18-26]

In the weak coupling approximation superfluid ³He is described in terms of *p*-wave BCS theory. In practice the onset of superfluidity modifies the pairing interaction, for example pairing mediated by the exchange of spin-fluctuations. These strong-coupling effects stabilize the bulk A phase at high pressures.[27] From details of the A-B phase boundary under confinement, and superfluid fractions at the transition, we show that strong-coupling effects persist to the lowest pressure.

The B phase of superfluid ³He is a time-reverse invariant odd-parity condensate of *p*-wave pairs, with all three components of the spin and orbital triplet present, and isotropic energy gap. The A phase is a chiral superfluid, in which only the equal spin pairs form with a common orbital angular momentum ***l*** vector. Near a wall, gap-distortions arise from de-pairing due to surface scattering; the A phase orients the ***l*** vector normal to the surface, minimizing gap suppression[28,29]. This weaker gap suppression favors the A-phase and is responsible for the profound influence of confinement. The relevant length scale over which gap distortion occurs is set by the pressure dependent Cooper pair diameter, $\xi_0 = \hbar v_F / 2\pi k_B T_c$ where $v_F$ is the Fermi velocity, $T_c$ is the bulk superfluid transition temperature, $\hbar$ and $k_B$ are Planck's and Boltzmann's constants. When two surfaces are separated by a comparable distance *D*, $D/\xi_0 \cong 10 - 20$, the distortion of the order parameter at the surface influences the entire sample, particularly approaching $T_c$, promoting the A phase over the B phase.[30]

Following Levitin et al.[17] we adopt the temperature dependent coherence length $\xi_\Delta(T) = \hbar v_F / ((\sqrt{10})\Delta_B(T))$, such that $\xi_\Delta(0) = 1.13\, \xi_0$, and $\xi_\Delta(T)$ tends to the Ginzburg-Landau (GL) result as $T \to T_c$. The predicted equilibrium AB transition temperature, $T_{AB}$, for slabs of different thickness is given by a universal value of $D/\xi_\Delta(T_{AB})$, which increases with pressure due to strong coupling. [17,31-33]

Until recently investigations of ³He under regular confinement were limited to arrays of plates and capillaries[34-36], or studies of saturated films.[37-39] Previously we confined ³He to a single 680 nm tall nanofluidic cavity, and used nuclear magnetic resonance (NMR) to map the phase diagram.

We found the B phase to be completely excluded at low pressure and observed the A phase between the normal state and the B-phase above 3 bar.[17,40] This contrasts with the bulk, where the A phase is confined to a region of high pressure (P≥ 21 bar) and high temperature (T≥ 1.9 mK). Guided by the results of previous work[17], for this experiment we chose a weaker confinement, with cavity height 1080 nm, corresponding to $D/\xi_\Delta(T=0, p=0) = 12$, such that an AB transition was expected to occur at all pressures.

A consequence of moderate confinement is the prediction of a novel spatially modulated, or "stripe phase", intervening between the A and B phases[30,41], that breaks translational symmetry in the plane of the slab, and is composed of alternating regions of degenerate B phase domains of different orientations. It is an analogue of Fulde-Ferrel-Larkin-Ovchinnikov phases,[42,43] long sought after in superconductors[44,45] and fermionic ultra-cold atom systems.[46-48] In $^3$He the driving mechanism for the spatially modulated phase is the negative surface energy of domain walls under confinement, that allow domains of degenerate B-phase "quantum vacua" to spontaneously appear. Recently the existence, location and stability of the putative stripe phase have been shown to depend crucially on details of the strong coupling parameters.[41] Here we argue that the presence of such a stripe phase, even if it is not the lowest free energy state, can promote the resonant tunneling mechanism for B-phase nucleation.[26] Such a mechanism is intrinsic without invoking the need for nucleation to be seeded extrinsically by neutrons or cosmic rays.[18-24]

## Results:

**Experiment details:**

The $^3$He was confined to a 1.08 µm deep cavity micro-machined in 1 mm thick silicon, capped with 1 mm thick sodium doped glass, anodically bonded[49] to the silicon. The cavity is shown in Fig. 1 a,b and construction details are provided elsewhere[50] and in Supplemental Fig. 1,2. Under pressure, the cavity distorts by 180 nm at 5.6 bar as found by finite element modeling and depicted in Fig. 1c, d and Supplemental Fig. 3 (see also Methods). The bowing of the cell plays an important role in our study of B-phase nucleation. For the measurements described here, the surfaces were coated with a 30 μmole.m$^{-2}$ coverage of $^4$He to eliminate solid $^3$He from the surface and allow direct comparison with the earlier NMR experiment, with diffuse scattering[51].

**Measurables:** We measure the resonant frequency, $f$, and the Quality factor, $Q$, of the torsion pendulum as a function of temperature. The $^3$He in the cavity is fully coupled to the torsion pendulum above $T_c$. The superfluid fraction is determined from the increase in resonant frequency (after subtracting the temperature dependent background of the empty oscillator), arising from the decoupling of the superfluid below $T_c$. The dissipation ($Q^{-1}$) was calculated by relating the observed amplitude at resonance to the drive voltage (after calibration at a fixed drive). Temperatures were measured using a melting curve thermometer[52,53] and then related to the frequency shift and dissipation of a quartz fork, immersed in the same heat exchanger as the $^3$He in the torsion pendulum.[54] Below we express temperatures in units of the bulk $T_c$, measured *in situ* by the fork. $T_c$ = 0.9-1.6 mK at p = 0.1-5.6 bar. The suppression of $T_c$ in the slab due to confinement was less than 0.01$T_c$, and is not discussed further.

**Superfluid fraction:** The superfluid fractions measured while cooling (blue circles) and warming (red open triangles) for five pressures are plotted in Fig. 2. The data highlighted by circles contains the signature of the AB transition shown in detail in Fig. 3. The superfluid fraction (dashed line) of the bulk B phase at each pressure[55] is essentially indistinguishable from that measured under confinement.

**Transition between A and B phases:** The region near the first order AB transition is the focus of this paper. In summary we study: (i) the influence of the confinement on the equilibrium transition temperature, $T_{BA}$ observed on warming, with the transition broadened by pressure-induced bowing of the cavity (Fig. 4a); (ii) the supercooling of the transition from A to B, $T_{AB}$, and its pressure dependence (Fig. 4a); (iii) the ratio of the superfluid densities at this transition (Table 1, Fig 4b). We describe the systematics of these data in turn, followed by a discussion.

**Warming transition:** The warming transition $T_{BA}$, determines the thermodynamic transition temperature, as confirmed by the lack of history dependence of this feature (Fig. 5a). However, this transition exhibits a finite width, because of the smooth variation in height across the annular cell's cross section, arising from pressure-induced bowing, as in ref. 17. The cross section of the annular region under pressure is shown in Fig. 1c,d and supplemental Fig. 3, (also Methods).

At each pressure we identify the start $T_{BA}^{lower}$ and end temperature $T_{BA}^{upper}$ of the transition (see details in Supplementary note 1) determining the transition's width, $\Delta T/T_c = (T_{BA}^{upper} - T_{BA}^{lower})/T_c$. The increase in transition width with pressure, (Table 1, Fig. 4a), is consistent with the calculated bowing. $T_{BA}/T_c$ increases with increasing pressure, driven by the decrease of the zero temperature coherence length; a comparison with theoretical prediction is made later.

**Supercooling:** The transition at $T_{AB}$ while cooling (blue circles Fig. 3a-e) is indicated by a small but abrupt jump in superfluid fraction (Fig. 4b and Table 1). The apparent width (in temperature) of this transition is limited by the cooling rate $\sim$ 10 μK.hr$^{-1}$ and the oscillator decay time ($\sim$1000 s). The supercooling of the A phase $\delta T/T_c = (T_{BA}^{upper} - T_{AB})/T_c$ is shown in Table 1. It is measured from the upper temperature (completion) of the B→A transition, since this corresponds to maximum cavity height, where the B phase is nucleated on cooling. The supercooling is extremely small in comparison with that observed in bulk, and exhibits a non-monotonic pressure dependence, Fig. 4a.

**Absence of Pinning:** The finite width of the BA transition observed on warming, attributed to cell bowing, implies that during this transition the AB interfaces in the cell are positioned such that $D/\xi(T_{AB})$ is at the critical value (see Supplementary note 1). The surfaces of the cavity were polished in an attempt to eliminate pinning. To confirm the absence of pinning the experiment was warmed partially into the B→A transition region and then re-cooled, Fig. 5. The data clearly shows no hysteresis between warming and cooling and therefore no pinning. Warming to just above $T_{BA}^{upper}$ and subsequent cooling well reproduces the "supercooled" trajectory. This behavior contrasts to the data obtained in the 680 nm cavity that showed hysteresis associated with pinning of the A-B interface at scratches on the glass surface of that cavity.[40] The supercooling and warming transitions are depicted schematically in Fig. 5 b,c.

**Superfluid density and dissipation:** The ratio of the superfluid density in the A phase and the corresponding value in the B phase is precisely determined at B phase nucleation, $T_{AB}/T_c$, Table 1. We also note that the dissipation is greater in the B phase (Figs 3a-e) than in the A phase. In the bulk, the reverse is true.[56] We observe that the dissipation excess at $T_{AB}$ is pressure dependent and decreases as the pressure increases. This may be indicative of a contribution to the dissipation from surface Andreev bound states in the B-phase, confined within distance $\sim \xi_0$ from each wall, where $\xi_0$ decreases with pressure.[57]

## Discussion:

In bulk, the conventional homogeneous nucleation theory predicts the lifetime of the supercooled A phase to exceed the age of the universe[58]. The nature of the mechanism for the nucleation of the B phase remains a matter of debate with several competing scenarios (Baked Alaska[21,23] Kibble-Zurek[24], Q balls[25], and resonant tunneling[26]). Several of these scenarios rely on an extrinsic mechanism in which local heating of the superfluid is caused by the energy deposited by an incoming particle (*e.g.* neutron or cosmic-ray).

Under confinement in our thin slab geometry we observe only very small supercooling, less than $0.03T_c$. By contrast, in bulk with clean surfaces, the observed lifetime of the supercooled A state is exceedingly long at temperatures near the equilibrium AB transition (*ie* the B phase does not nucleate). Only by cooling to very low temperatures ($\cong 0.25T_c$, far below the equilibrium AB transition), is the transition to the B phase observed while holding the temperature fixed for several hours. Our experimental practice involves cooling at a steady rate through the supercooled state for several hours; if the "lifetime" of the supercooled A phase under confinement were the typical bulk value of an hour, we would expect a supercooling of $\cong 0.4T_c$.[18]

The critical radius of a bulk B phase nucleation bubble at $T = 0$, $R_0$, is inferred from experiment to be of the order of 0.5 μm at high pressures, and in zero magnetic field scales approximately as $R(T) = R_0(1 - T/T_c)^{1/2}/(1 - T/T_{AB})$, diverging at the equilibrium $T_{AB}$, where the free energy difference between the two phases vanishes[18]. We infer from measurements of the surface energy at $p = 0$,[59] and other thermodynamic data, that $R_0$ is comparable at low pressures. Since, under confinement, supercooling is small, *R(T)* is much greater than $R_0$ and therefore *D*; the relevant nucleation volume is a disc of height *D* and radius *R*, ruling out homogeneous nucleation due to its macroscopic size.

The sample geometry of our cavity is particularly well suited to studies of B phase nucleation. Due to the influence of confinement on the relative stability of the A and B phases, the B phase nucleates in the most bowed region of the cell, as depicted in Fig. 5b. The edges of the annular confinement region thus act as a "bottle" isolating the interior region, where B phase nucleation occurs, from any B phase otherwise present (for example in the fill line). This is functionally equivalent to the small 0.6 T NdFe permanent magnets that were used in experiments at Stanford.[18] There the locally strong magnetic field stabilized the A phase creating a "valve" to isolate the helium under study in quartz tubes from bulk $^3$He-B nucleated in a sinter or other poorly characterized material. Our cell's surface roughness is well characterized and much smoother than the quoted value (<10 nm) in the Stanford experiments.[18] This is relevant since surface roughness is also implicated in the nucleation process.

The smallness of the supercooling and the smallness of the sample volume, strongly suggest that the B-phase nucleation we observe under confinement is a new intrinsic phenomenon. We note that the AB transition in aerogel, by contrast, shows large supercooling[19,20] which suggests that surfaces by themselves are not important. On the other hand, the Tye-Wohns mechanism[26] invokes the presence of an intermediate state between the A phase and the B phase which aids nucleation via resonant tunneling. Under confinement, a natural intermediate state exists: the stripe phase. However strong coupling corrections are predicted to strongly influence the stability of the equilibrium phase[41]. The two scenarios are therefore: (i) over some region of the phase diagram the striped phase is stable, and the A phase makes a first order transition into the striped phase which continuously evolves into the B-phase; (ii) the striped phase is not stable but provides the required nearby set of quantum vacua to nucleate the B phase via the resonant tunneling mechanism.

The calculated phase diagram[41] is shown in Fig. 6, determined from GL theory, which strictly applies in the $T \to T_c$ limit, and based on the strong coupling $\beta$ parameters proposed by Choi *et al.*.[60] This calculation applies a correction to standard GL theory, which takes into account a linear scaling in temperature of the strong coupling correction to the weak coupling *β* parameters, and successfully reproduces the experimental bulk phase diagram at high pressures. These strong coupling parameters do not extrapolate to the weak coupling limit at zero pressure, consistent with earlier work which concluded that strong coupling corrections continue to play a role at zero pressure.[61] With these parameters the stripe phase is restricted to a thin sliver of the pressure-temperature phase diagram between 1 and 3 bar,[41] (Fig. 6) in contrast to a significant wedge predicted on the basis of GL theory with *β* parameters derived from theory (Supplemental Fig. 4).[41,62] The region of stability of the stripe phase, if any, is thus an open question. In the present experiment the observed minimum in the extent of the supercooling of the A phase (Fig. 4) is well aligned with the predicted region shown. It is suggestive that the minimum in supercooling may arise from the virtual presence of the striped phase to mediate the nucleation of the B phase. To validate the resonant tunneling model one would have to quickly traverse the region where the striped phase is marginally stable and examine the statistics of nucleation. We were unable to carry out rapid cooling strategies or measurements at high pressures due to technical factors.

The growth of the A phase from the B phase (Fig. 5), is not subject to any nucleation barrier. There are small features at the side walls of the cavity, where the A phase (favored because of confinement) is still present even after cooling deep below the observed completion of the A→B transition (Supplemental Fig. 2). We follow a procedure (see supplemental note 1) to determine a best fit value at the thermodynamic B→A transition for $D/\xi_\Delta(T)$ at each pressure and plot the best fit for the measured $\rho_s/\rho$ vs. $T/T_c$ in Fig. 3 as the solid line. The locations in $T/T_c$ of the start of the B→A transition ($T_{BA}^{lower}/T_c$) and the end of the B→A transition ($T_{BA}^{upper}/T_c$) along with $D/\xi_\Delta(T)$ at each pressure are also listed in Table 1.

**Phase Diagram:** The measured phase diagram, in comparison with the predicted phase diagram[41] for this cavity height, is shown in Fig. 6. To test the universality of $D/\xi_\Delta(T_{BA})$ we also compare the present measurements on a 1080 nm cavity, using the superfluid density, and the previous NMR experiment on a 680 nm cavity (Fig 6b-d).[17,40] There is good agreement of $D/\xi_\Delta(T_{BA})$ between the two experiments (Fig. 6b). In weak coupling theory $D/\xi_\Delta(T_{AB})$ is pressure independent and depends only weakly on specularity of surface scattering[63-65], (Fig 6b). Thus we confirm that strong coupling corrections play a role at all pressures. The most recent theory, Fig. 6, including a proposed temperature dependence of the strong coupling corrections,[41,66] still shows significant discrepancies with the data, emphasizing the uncertainty in our current knowledge of the strong coupling parameters at low pressure. However the small shift between the two cavity heights' data sets is consistent with the shift between the theoretical lines, which show that the $D/\xi_\Delta(T_{AB})$ phase boundary is no longer universal, due to the proposed temperature dependence of the strong coupling corrections. This plot also highlights the sensitivity of the putative equilibrium stripe phase to the precise details of pressure dependent strong coupling parameters and illustrates the pressure window over which a "nearby" stripe phase may play a role in B phase nucleation via the resonant tunneling mechanism.

Due to bowing, the A-B interface will be oriented azimuthally, and hence parallel to the flow. Thus the putative striped phase would be located at this interface, parallel to the flow with expected minimal effect on the superfluid fraction in our experiment. The present set-up cannot rule out a thermodynamically stable stripe phase,

The observed superfluid density of the A phase at $T_{AB}$ is greater than that of the B phase (for $\rho_{s\perp}^A/\rho_s^B(T_{AB})$ see Table 1, Figs. 3, 4b). The component of the superfluid density tensor assayed is $\rho_{s\perp}^A$ [67] since the $l$ vector is oriented perpendicular to the surfaces.[68] $\rho_{s\perp}^A/\rho_s^B(T_{AB})$ decreases as the pressure is lowered, Table 1, Fig. 4b. Although the B phase gap is subject to a planar distortion due to confinement, the superfluid density is sensitive to the parallel component of the gap, which should be close to the bulk isotropic value.[55] Detailed calculations of the superfluid density under confinement are however required. Since the ratio of superfluid fractions is unity in the weak coupling limit and our data always shows that $\rho_s^A/\rho_s^B > 1$, this measurement provides further support for the presence of strong coupling even at the lowest pressures.[69]

In conclusion we have made the first torsional pendulum study of superfluid $^3$He confined in a single micron sized cavity. The ease with which the B phase nucleates provides clear evidence for an intrinsic nucleation mechanism at the first order A→B transition under such confinement. We established, via the scaling of the equilibrium AB transition temperature with cavity height, that strong coupling effects at low pressures are stronger than currently believed. This has implications for the stability of the stripe phase, which was not directly observed in this experiment. However the observations of B phase nucleation are consistent with the resonant tunneling scenario where we propose that, even if not stable, the stripe phase can provide the intermediate metastable state to mediate the transition. Since the energetics of the stripe phase appear be particularly sensitive to choice of cavity height and pressure, the detailed systematics of B phase nucleation, in response to temperature and pressure changes for different cavity height, should provide further test of our hypothesis. Nanofluidic cells also offer the potential to control and manipulate the A-B interface via stepped size modulation.

It is believed to be likely that first order phase transitions occurred in the early universe, may explain matter-antimatter asymmetry, and are important in inflation scenarios. According to standard nucleation theory the supercooled A phase lifetime in bulk superfluid should greatly exceed the lifetime of the universe, and this has prompted proposals for a number of extrinsic nucleation mechanisms. The fact that supercooling is virtually eliminated under confinement is striking. The notion that confinement significantly modifies the energy landscape, creating false vacua that promote B phase nucleation, via resonant tunneling, merits further study in the laboratory, which may impact on our understanding of these central questions in cosmology.

## Methods:

**Cell Construction:**

The 14 mm silicon disk that comprises the micromachined chamber to contain the $^3$He was fabricated at Cornell's Nanofabrication facility.[50] The process flow is shown in Fig. 1 of the supplementary materials. After patterning of the silicon a matching octagonal piece of highly polished Sodium doped glass (Hoya SD-2) was bonded to the silicon. The so-constructed cavity comprises the head of the torsion pendulum, and was mounted to the coin silver torsion rod using a special alignment jig and Tra-Con epoxy. The free volume in the torsion rod was minimized by using a quartz tube of 320 μm outer diameter 100 μm inner diameter that also served to exclude the epoxy (because of its high viscosity) from flowing into the cavity.

**Characterization of cavity geometry:**

The fabricated cavity height (1080 nm) is maintained only at the bonded points of attachment. The step height on the cavity was measured prior to bonding with a Tencor P10 profilometer and found to be 1080 nm. Roughness of the silicon cavity surface was measured using an Atomic Force Microscope (AFM) to be less than 0.1 nm rms. The same measurement for the glass surface after polishing was seen to be less than 1 nm rms.

Finite element analysis (Fig. 1) and Supplemental Fig. 3 reveals the extent to which walls of the rectangular cross section nanofluidic cavity bow under pressure. We modelled the bowing using finite element methods (COMSOL) and materials properties from standard tables that produce an expected bow (at maximum) of 30 nm.bar$^{-1}$. Measurements on a different cavity made from the same glass and silicon and having geometry suitable for NMR investigations yield a best fit of 32 nm.bar$^{-1}$ bowing. The fits shown in Fig. 3 use the 31 nm.bar$^{-1}$ figure and the error bars for $D/\xi_\Delta(T_{BA}^{lower})$, $T_{BA}^{lower}/T_c$, and the resulting $\Delta T/T_c$ in Table 1 and figures, were obtained taking into account the difference between the calculated and experimental figures. Error bars for $T_{AB}$, $T_{BA}^{lower}/T_c$, and the resulting $\delta T/T_c$ are determined from the noise in the experimental data.

The edge of the annular cell cavity retains features due to the fabrication process where small regions of a few μm width of A phase likely persist well below $T_{AB}$. These features are illustrated in Supplementary Fig. 2.

**Torsion Pendulum:**

Coin silver (90% silver and 10% copper) was chosen as the alloy for the torsion pendulum because this material provides a high Q at low temperatures and has a relatively small temperature dependent frequency background.[70] As with our usual practice we drove and detected the pendulum motion electrostatically, keeping it close to resonance using a digital phase locked loop. The resonant frequency of the mode we excited the pendulum at was about 1330 Hz and the quality factor was ~$1.5 \times 10^6$ at mK temperatures. Below the superfluid transition, the superfluid fraction of the fluid in the cavity decouples from the pendulum, and a decrease in its period is observed. The ratio of the moment of inertia of the fluid in the cavity to the moment of inertia of pendulum head was ~6 parts in $10^6$. The frequency noise is about 2-4 parts in $10^9$ giving us ample resolution to resolve the superfluid fraction $\rho_s/\rho$. Achieving this degree of frequency stability required a compromise between driving the pendulum at a large enough amplitude (to increase signal above ambient vibrational noise) and non-linearity of the pendulum's torsional mode.

The superfluid fraction was determined from the frequency of the torsion pendulum, $f(T)$, through the following expression:

$$\frac{\rho_s}{\rho} = \frac{\Delta f(T)}{\Delta f_{fluid}} \qquad (1)$$

where $\Delta f(T)$ is the difference between the measured frequency $f$ and the frequency of the pendulum if all the fluid is fully locked $f_{full}$.. At $T_c$ the viscous penetration depth of $^3$He is about 0.5 mm, many orders of magnitude larger than the distance between the plates. We can assume that for temperature slightly above $T_c$ to $T_c$ all the fluid is fully coupled to the pendulum and $f_{full} = f$. For values of $f_{full}$ below the superfluid transition we extrapolate from the normal state values. $\Delta f_{fluid}$ is equal to the difference in the pendulum frequency between empty and filled . $\Delta f_{fluid}$ should scale linearly with the density of the fluid. When the sample is pressurized, the axis of the pendulum distorts slightly. To avoid any error due to this,  Below 1 K, $^4$He is nearly 100 % superfluid, so any difference between the filled cell and empty cell frequency below 1 K will be only due to torsion rod distortion. By comparing empty cell data and data for $^4$He at $T < 1$ K we determine that this effect was responsible for a frequency shift of 1.78 mHz for 3 bar of pressure. After the effect of torsion rod distortion is accounted for, we observe that the difference for the torsion pendulum frequency between fully coupled fluid just above $T$ and the empty cell frequency is 7 mHz. Using the appropriate values for the density of $^4$He at these temperatures and pressure, we determine that the expression for the frequency shift due to the fully coupled normal fluid is:

$$\Delta P_{fluid} = 46.05 \text{ [mHz.cm}^3\text{.g}^{-1}\text{]} \times \rho \text{ [g.cm}^{-3}\text{]}$$

where $\rho$ is the density of $^3$He at the superfluid transition.

# End Notes:

**Acknowledgements:**

We acknowledge input from H. Tye, M. Perelstein, E. Mueller, J.A. Sauls, J.J. Wiman, Hao Wu, and A. Casey. We also acknowledge the assistance of R. DeAlba with imaging of the silicon surface depicted in Supplemental Fig.2. This work was supported at Cornell by the NSF under DMR-1202991 and in London by the EPSRC under EP/J022004/1, and the European Microkelvin Platform.

**Author contributions:**

Experimental work and analysis was principally carried out by N.Z. assisted by T.S.A. with further support from E.N.S. and J.M.P. R.G.B. and N.Z. established the nano fabrication protocols, and X.R. and R.G.B. carried out the finite element modelling. L.L. analyzed much of the data for comparison of the 680 nm and 1080 nm data sets, and N.Z. and L.L. shared in much of the eventual data analysis. J.M.P. supervised the work and J.M.P. and J.S. had leading roles in formulating the research and writing this paper.

**Competing financial interests:**

The authors declare no competing financial interests.

**Data Availability:**

The data that supports this study is available through Cornell University e-commons data repository at http://hdl.handle.net/1813/44729

Figure 1: Torsion pendulum head

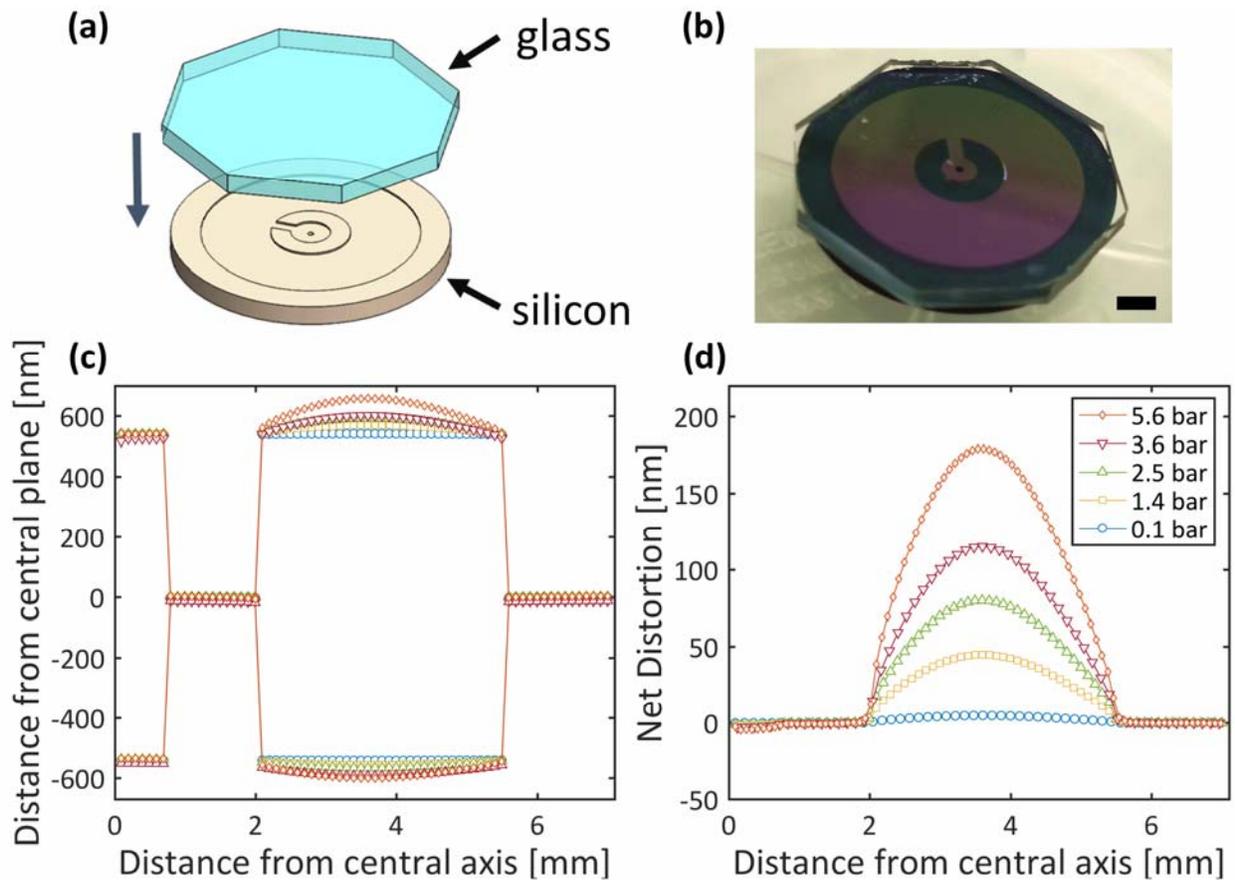

Legend:

a) Schematic of the glass-silicon head. The 1 mm thick × 14 mm diameter silicon was patterned with a 1.08 μm tall × 11 mm outer diameter / 4 mm inner diameter cavity before the octagonal glass lid was anodically bonded to it. The annular cavity is connected to the central fill line through a 1.25 mm long × 0.6 mm wide radial channel that opens into a 1.5 mm diameter central hub.
b) The bonded cavity prior to mounting on the torsion pendulum. Scale bar is 2 mm.
c) The cross section of the cavity under pressure.
d) The calculated net bowing of the cavity at the five experimental pressures investigated. (c) and (d) were calculated using finite element methods.

Figure 2: Temperature dependence of the superfluid fraction

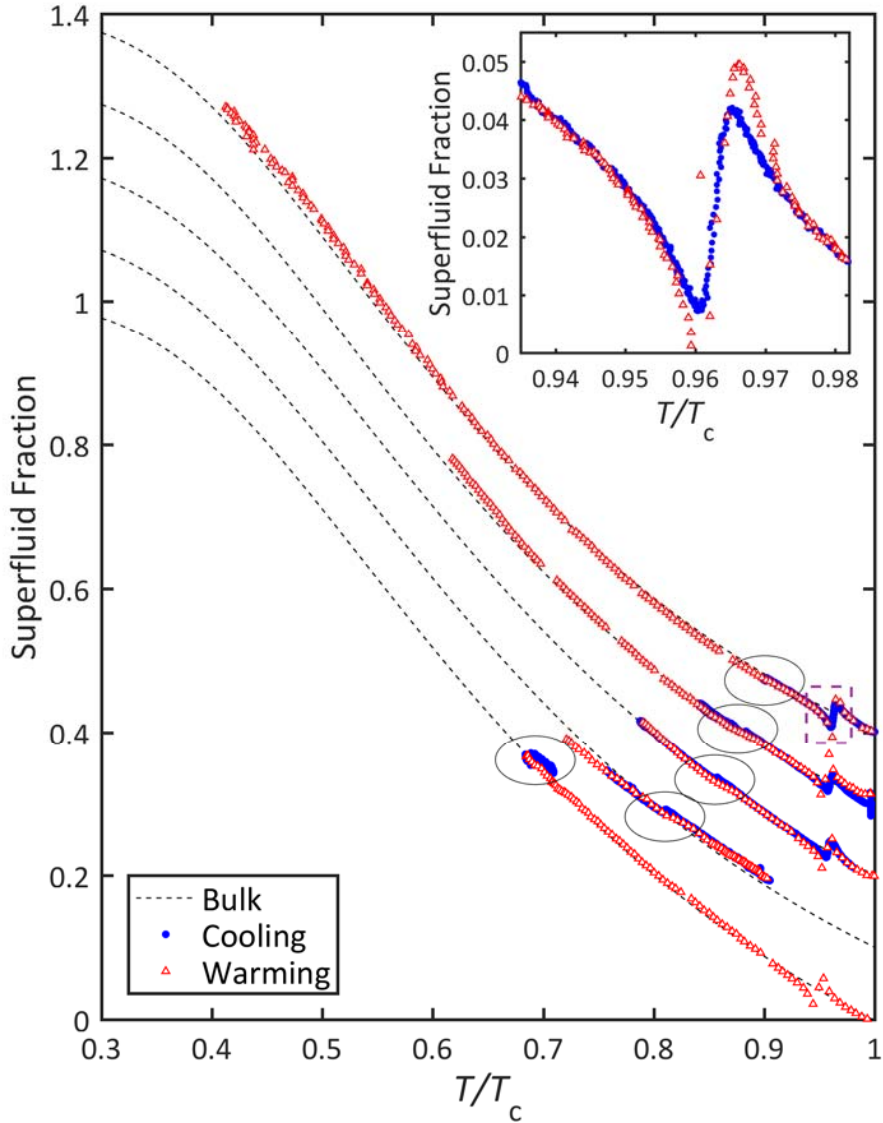

Legend:

Superfluid fraction (offset by 0.1 for clarity) measured at 0.1 (bottom), 1.4, 2.5, 3.6 and 5.6 bar (top) while warming and cooling as a function of temperature measured in units of bulk superfluid transition temperature $T_c$. Encircled are the locations of the A to B and B to A transitions, shown in detail in Fig. 3. Dashed lines show the superfluid density of the bulk B phase[55]. The antisymmetric signature just below $T_c$ arises from a mode crossing of a Helmholtz resonance with the torsional mode (marked with a dashed box and highlighted in inset). The crossing is narrower after warming from the B phase into the A phase than after cooling from the normal fluid into the A phase.

Figure 3: Superfluid fraction and Dissipation at A-B transition while warming and cooling.

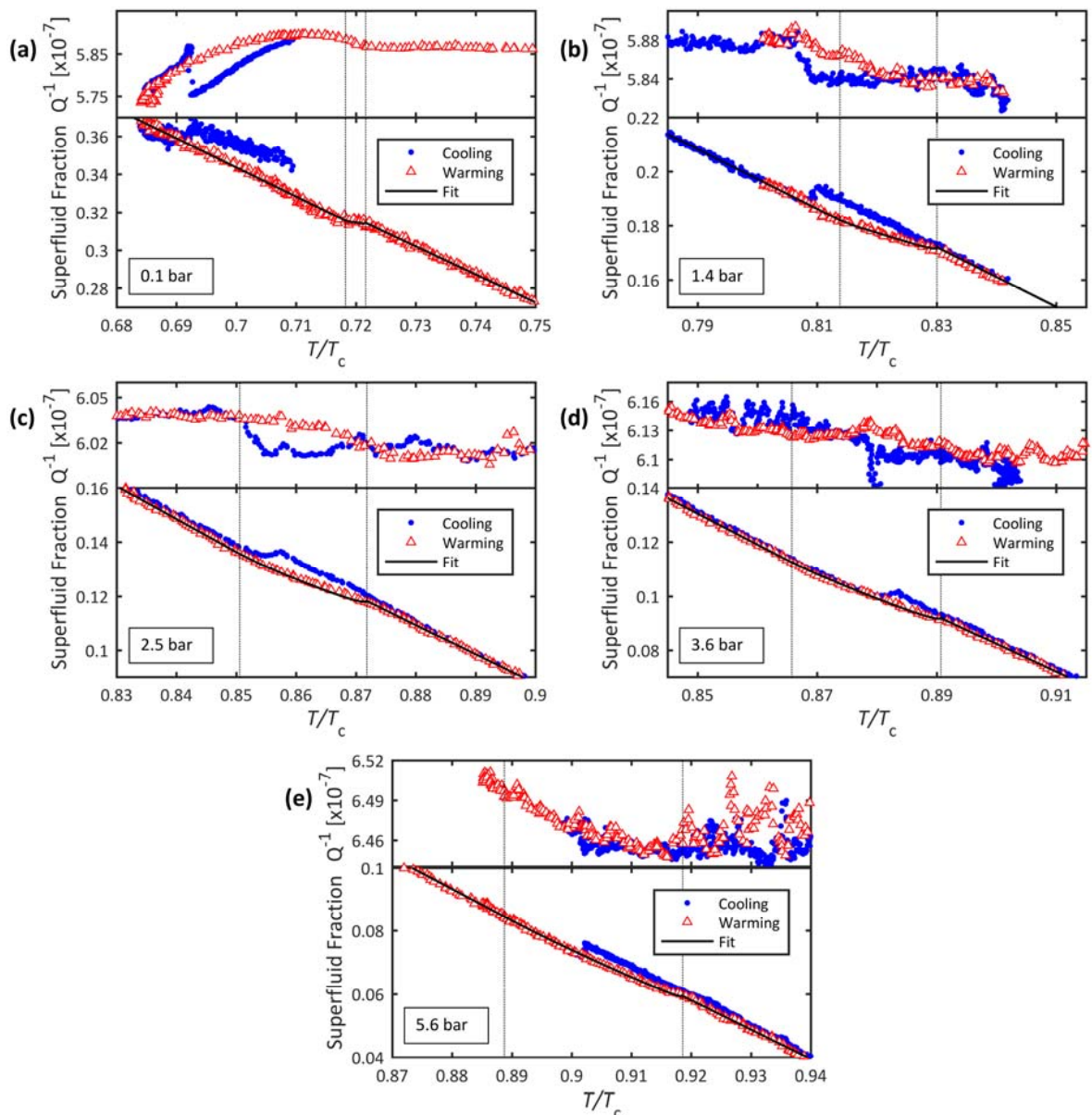

Legend: The dotted vertical lines mark the onset $T_{BA}^{lower}/T_c$ (left) and completion $T_{BA}^{upper}/T_c$ (right) of the B to A transition on warming. This gradual transition is associated with the pressure-induced cavity height distribution; solid black lines show fit to the data assuming the transition at constant reduced thickness $D/\xi_\Delta(T_{AB})$. On cooling (blue circles) the B phase nucleates abruptly from the supercooled A phase. At 2.5-5.6 bar the supercooling is smaller than the width of the warming transition, and the jump corresponds to a transition from pure A phase into spatially-separated A/B phase coexistence. The dissipation increases across the A to B transition, particularly at low pressures.

Figure 4: Pressure dependence of the measured properties of the A-B transition.

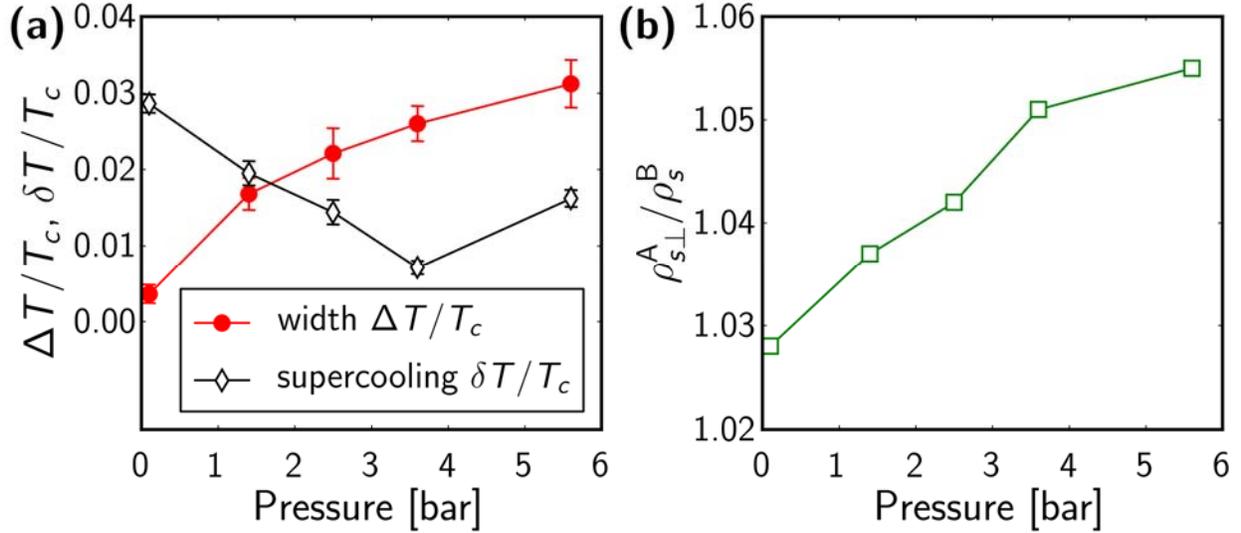

Legend:

a) The width of the B to A transition observed on warming (due to the pressure-induced bowing of the cavity) is compared to the extent of supercooling, which shows a non-monotonic behavior, suggestive of the stripe-phase-mediated resonance tunneling mechanism of the B phase nucleation.
b) The ratio $\rho_{s\perp}^A/\rho_s^B$ of superfluid fractions in the A and B phases near the transition; the departure from unity indicates the presence of the strong coupling effects down to low pressure.

Figure 5: Traversal of the A B transition

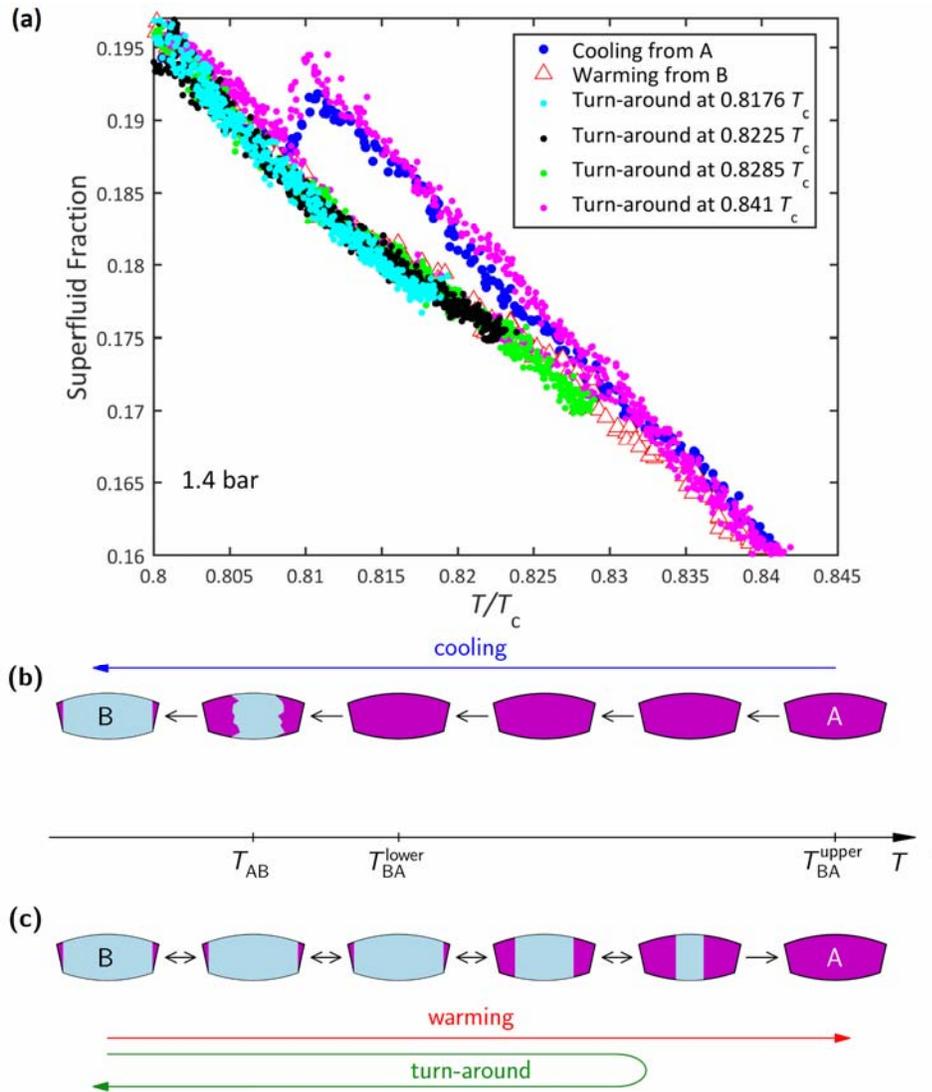

Legend:

(a) Superfluid fraction on different trajectories through the A-B transition region at 1.4 bar. (b,c) Schematic depiction of the distribution of the A and B phases in the cavity in the presence of bowing. (b) The A phase supercools down to $T_{AB}$, where the sample converts into the B phase at once. For 1.4 and 0.1 bar $T_{AB}$ occurs below $T_{BA}^{lower}$. At 2.5, 3.6 and 5.6 bar, $T_{AB}$ occurs above $T_{BA}^{lower}$ and the transformation from A to B phase continues on cooling below $T_{AB}$ till the B phase fills the annular cavity at $T_{BA}^{lower}$ with the exception of the tapered edges (see supplemental Fig. 2) which remain in the A phase due to strong confinement. (c) These edges serve as seeds of the A phase at the gradual B to A transition on warming via the A/B coexistence state. If the sample is cooled again from the coexistence state, the A to B transition also occurs gradually, following the route of the warming transition, indicating that on warming the equilibrium transition is observed.

Figure 6: Phase diagram of superfluid $^3$He confined in slab geometry.

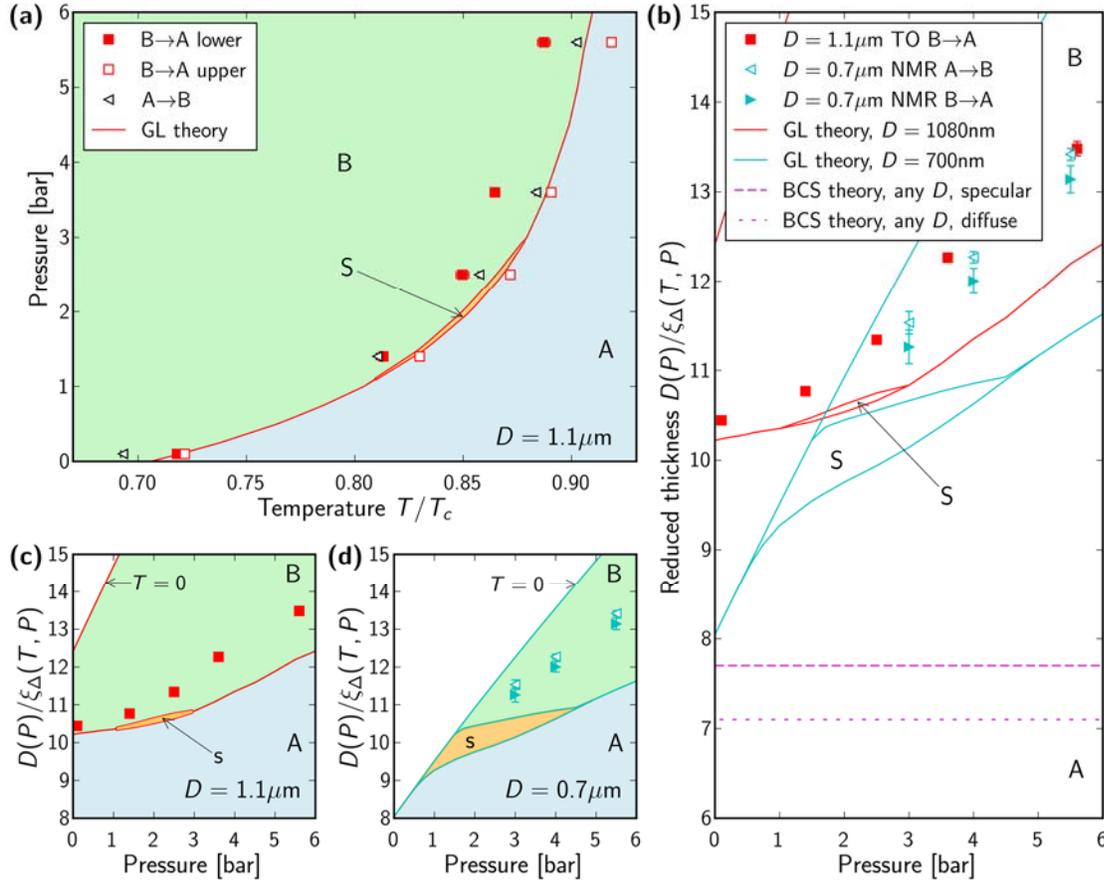

Legend: (a) The A-B transition in the $D=1.1$ μm slab in the temperature-pressure plane. The start, $T_{BA}^{lower}$, and completion, $T_{BA}^{upper}$, (filled/open red squares) of the B to A transition on warming correspond to the equilibrium A-B transition in the thinnest (fixed $D=1080$ nm near the edge) and thickest (pressure-dependent $D$) parts of the cavity respectively. On cooling the A phase supercools down to $T_{AB}$ (black triangles). Solid red lines and blue/orange/green areas show the A/stripe/B phase diagram predicted by the Ginzburg-Landau (GL) theory[41] with experimentally determined strong coupling parameters for $D=1080$ nm, therefore directly comparable to $T_{BA}^{lower}$. The A phase is observed in the predicted region of stability of the stripe phase (orange), demonstrating the inaccuracy of the strong coupling parameters used for the calculation[41]. (b) The reduced thickness $D/\xi_\Delta$ representation of the phase diagram allows to compare this work (in this representation the $T_{BA}^{lower}$ and $T_{BA}^{upper}$ features coincide) with the NMR experiment on a $D=0.7$ μm slab[17,40] (due to hysteresis the equilibrium transition was not observed, here the data on warming and cooling are shown). The collapse of the two datasets demonstrates the universality of $D/\xi_\Delta$ at the A-B transition. These measurements strongly deviate from the prediction of the weak coupling (BCS) theory, shown here for diffusely and specularly scattering cavity walls[33]. Since the state of the boundaries does not play a major role in the location of the BCS A-B boundary, we compare the experiments with diffuse walls to the predictions of the GL theory only available for specular walls[41] (same in panel (a)). The phase boundaries derived within the GL theory for $D$=1080 nm and 700 nm (red and cyan lines; also shown separately in (c,d)) depart from the $D/\xi_\Delta$ collapse due to the temperature dependence of the strong coupling included in the theory[41]. A qualitatively similar shift is observed between the experimental datasets. The disagreement between the experiments and GL theory emphasizes the current limited understanding of the strong coupling parameters at low pressure that leaves the stability of the stripe phase uncertain.

Table 1: Measured Properties of the A-B Transition in a 1.08 μm slab.

| P [bar] | $\rho_{s\perp}^A/\rho_s^B(T_{AB})$ | $T_{AB}$ | $T_{BA}^{lower}/T_c$ | $T_{BA}^{upper}/T_c$ | $\delta T/T_c$ (supercooling) | $\Delta T/T_c$ (width) | $D/\xi_\Delta(T_{BA}^{lower})$ |
|---|---|---|---|---|---|---|---|
| 0.1 | 1.028 | 0.693 ± 0.001 | 0.718 ± 0.001 | 0.7216 ± 0.0006 | 0.0286 ± 0.0012 | 0.0036 ± 0.0012 | 10.428 ±0.008 |
| 1.4 | 1.037 | 0.8105 ± 0.0005 | 0.8132 ± 0.0015 | 0.8300 ± 0.0015 | 0.0195 ± 0.0016 | 0.0168 ± 0.0021 | 10.733 ±0.018 |
| 2.5 | 1.042 | 0.8574 ± 0.0005 | 0.8497 ± 0.0029 | 0.8718 ± 0.0015 | 0.0144 ± 0.0016 | 0.0221 ± 0.0033 | 11.28 ±0.03 |
| 3.6 | 1.051 | 0.8836 ± 0.0005 | 0.8647 ± 0.0022 | 0.8907 ± 0.0015 | 0.0071 ± 0.0016 | 0.026 ± 0.0027 | 12.175 ±0.045 |
| 5.6 | 1.055 | 0.9023 ± 0.0005 | 0.8873 ± 0.0029 | 0.9185 ± 0.001 | 0.0162 ± 0.0011 | 0.0312 ± 0.0031 | 13.32 ±0.08 |

Legend: The superfluid fraction ratio $\rho_{s\perp}^A/\rho_s^B(T_{AB})$ at the A-B transition, the temperature $T_{AB}/T_c$ of the A to B transition on cooling, together with the temperature $T_{BA}^{lower}/T_c$ of the start (lower end) of the B to A transition on warming and $T_{BA}^{upper}/T_c$ (upper end) where all the B phase is converted to the A phase are presented. The extent of supercooling $\delta T/T_c = (T_{BA}^{upper} - T_{AB})/T_c$ and the width of the B to A transition $\Delta T/T_c = (T_{BA}^{upper}/T_c - T_{BA}^{lower}/T_c)$ are derived. The best fit reduced thickness of the warming transition $\xi_\Delta(T)$ at $T_{BA}^{lower}$.

Supplemental Fig 1 Depiction of Silicon Fabrication Process

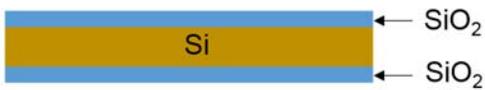
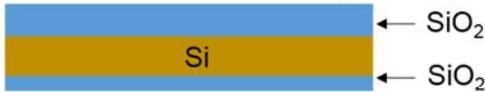
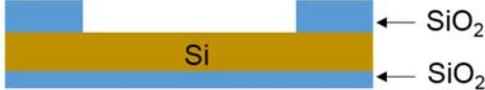
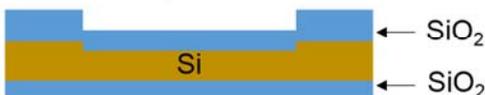
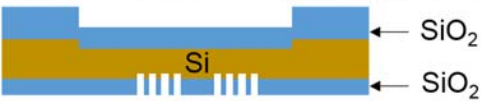
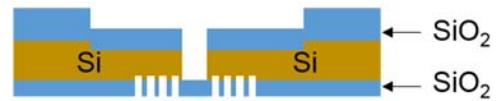
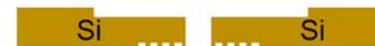
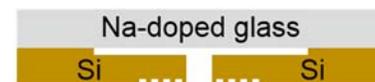
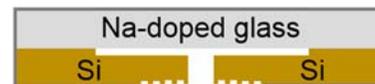

Legend:

Process flow for patterning of Silicon wafer. 1. Growth of thick oxide – as supplied from foundry. 2. Deposition of secondary oxide layer, followed by patterning, exposure and 3. dry/wet etch to strip oxide. 4 Further thermal oxidation to define cavity in silicon. 5. Pattern backside to define concentric circles for fill line attachment. 6. Reactive ion-etch to define fill line. 7. Strip off all oxide, clean and characterize surface 8. Anodic bonding step to complete cavity. 9. Deposit silver film by sputtering.

Supplemental Fig. 2 Detail at Cavity Edge

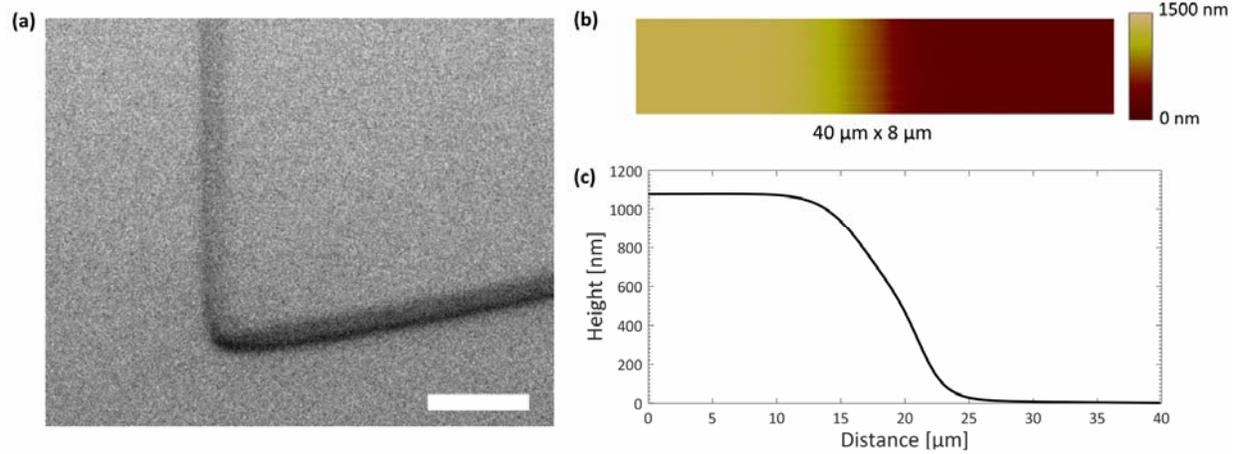

Legend:
  (a) Scanning electron micrograph showing the inner cavity wall and the edge of the central channel (the scale bar is 40 µm). The walls of the cavity are rounded due to the oxidation process with a lateral extend of approximately 10 µm. The A phase is stabilized well below the studied A-B transition in this region near the cavity boundary, since the cavity depth is less than 1080 nm there.
  (b) Atomic force microscope image of step at the outside cavity wall. Scanned region is 40 µm × 8 µm.
  (c) Line slice through the AFM scan showing clearly the profile of the cavity wall. Evident from this plot is the size of cavity of 1080 nm.

Supplemental Fig 3: Depiction of deflection of cell walls at 5.6 bar

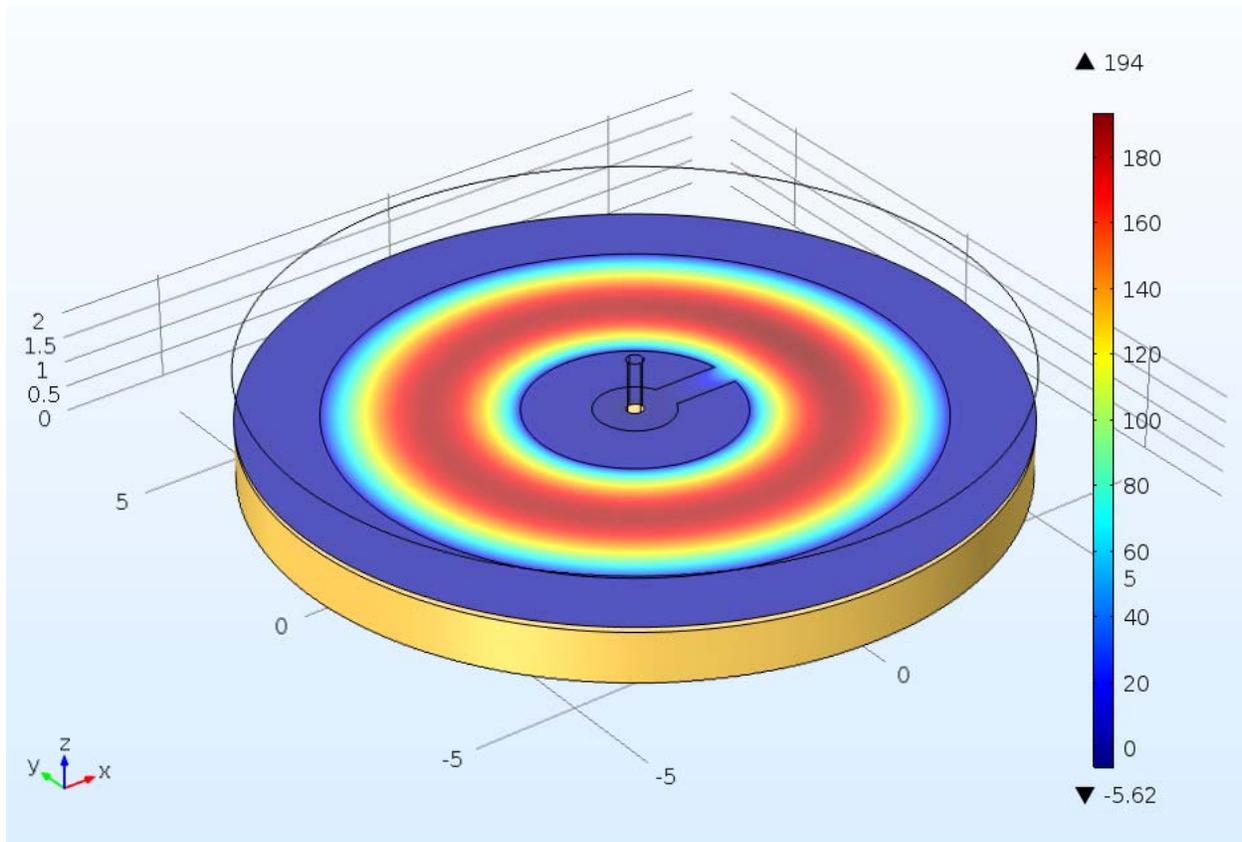

Legend:
Here we show the deflection of the cell walls at 5.6 bar computed using COMSOL. The central "hub" region has a minimal deflection and is expected to be in the A-phase when the annular region is in the B phase.

Supplemental Fig. 4 Phase diagram of superfluid $^3$He confined in slab geometry.

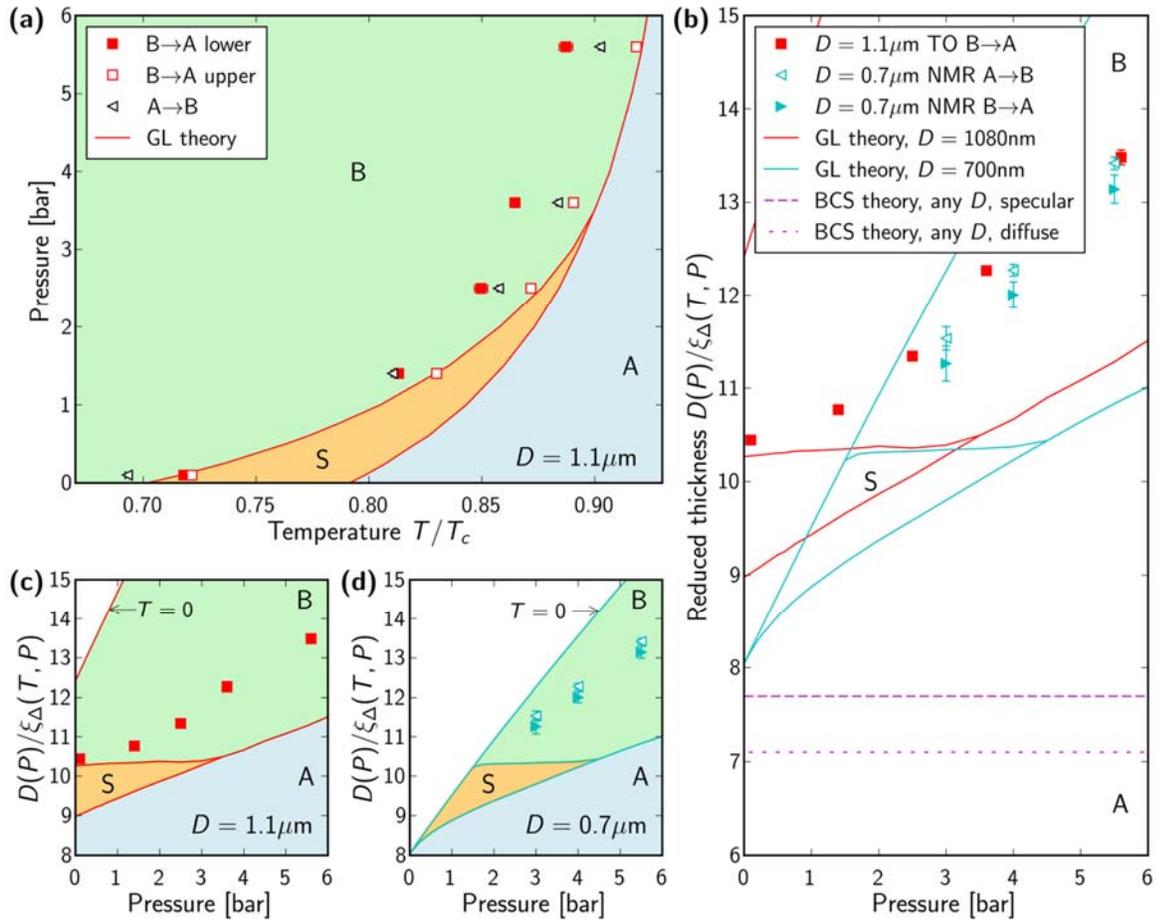

Legend: As described in Fig. 6 (main text) except that the predictions of the Ginzburg-Landau (GL) shows are derived using the strong coupling parameters calculated by Serene and Sauls [Supplemental Ref. [1]]. (a) The A-B transition in the $D$=1.1 μm slab in the temperature-pressure plane. (b) The reduced thickness $D/\xi_\Delta$ representation of the phase diagram used for comparison with the NMR experiment on a $D$=0.7 μm slab. The phase boundaries derived within the GL theory for $D$=1080 nm and 700 nm are shown with red and cyan lines; also shown separately in (c) and (d) together with the relevant experimental data.

---

[1] Sauls, J.A and Serene, J.W., Potential Scattering Models for the Quasiparticle Interactions in Liquid $^3$He, *Phys. Rev. B*, **24**, 183, (1981)

**Supplemental Note 1: Calculating the superfluid fraction through the B to A transition**

To fit for the best value of $D/\xi_\Delta (P, T_{BA}/T_c)$ at a given pressure we followed the procedure outlined below. We divide the annular channel into N equal radial elements.

1. We obtain linear fits for $\rho_s^B(T/T_c)$ and $\rho_s^A(T/T_c)$ near $T_{BA}/T_c$.

2. From the finite element modelling, we obtain a table of D(r, P) where r is the radius of a fluid element measured from the TO axis at any given pressure, $P$ using a value of 31 nm.bar$^{-1}$ for the maximum deflection.

3. We write the superfluid fraction

$$\frac{\rho_s^{Fit}}{\rho} = \frac{\sum_{i=1}^{N} \frac{\rho_s^{A\,(or)B}}{\rho}(T/T_c) \times 2\pi r_i^3 \times D(r_i) \times ((R_2-R_1)/N)}{\int_{R_1}^{R_2} 2\pi D(r) r^3 \, dr}.$$ (Supplemental Eq. 1) where

we have broken the cavity into $N = 350$ radial shells, each with radius $r_i$, ranging from $R_1$ = 2 mm to $R_2$ = 5.5 mm.

4. We select a trial value of $D/\xi_\Delta \left(P, \frac{T_{BA}}{T_c}\right) = \theta$. For any value of $T/T_c$, we determine if a particular element $r_i$ in S-1 has a value of $D(r_i)/\xi_\Delta(T/T_c) \geq \theta$. If the value is $\geq \theta$, then the element is taken to be in the B phase, if the value is $< \theta$, then it is taken to be in the A phase. We compute $\frac{\rho_s^{Fit}}{\rho}(\theta, T/T_c)$ for a given value of $\theta, T/T_c$.

5. We step $T/T_c$ through the region of interest.

6. At the end of steps 1-5 we can generate $\frac{\rho_s^{Fit}}{\rho}(\theta, T/T_c)$

7. The process is repeated for different values of $D/\xi_\Delta = \theta$

8. For a particular value of $D/\xi_\Delta = \theta$, we determine $T_{BA}^{upper}/T_c$, $T_{BA}^{lower}/T_c$. In particular, $T_{BA}^{upper}/T_c$ is very sensitive to small changes in our choice for $\theta$. However, $T_{BA}^{upper}/T_c$ and its variance can be directly determined from the data as the location at which superfluid fraction data on cooling and warming join. We compare the fit to measured values and select the best fit. This establishes best fit values for $D/\xi_\Delta$, $T_{BA}^{upper}/T_c$, $T_{BA}^{lower}/T_c$. Variances for the best fit for $D/\xi_\Delta$ are gauged when the fits clearly deviate from data. Values for $T_{BA}^{lower}/T_c$ and their variance follow from the best fit and variance of $D/\xi_\Delta$.

9. The process is repeated for maximum deflection values of 30 and 32 nm.bar⁻¹. Variation of these values affect the errors of $T_{BA}^{lower}/T_c$. These are added quadratically to the error found in step 8.